# Hide and seek in Slovakia: utilizing tracking code data to uncover untrustworthy website networks

Jozef Michal Mintal[1;2] [0000-0003-1651-7146], Michal Kalman[3], and Karol Fabián[1] [0000-0003-1424-0737]

[1] Department of Security Studies and UMB Data & Society Lab at CEKR, Matej Bel University, Banská Bystrica, Slovakia
`jozef.mintal@umb.sk`
[2] Center for Media, Data and Society at CEU Democracy Institute, Central European University, Budapest, Hungary
[3] Morione, Bratislava, Slovakia

**Abstract.** The proliferation of misleading or false information spread by untrustworthy websites has emerged as a significant concern on the public agenda in many countries, including Slovakia. Despite the influence ascribed to such websites, their transparency and accountability remain an issue in most cases, with published work on mapping the administrators and connections of untrustworthy websites remaining limited. This article contributes to this body of knowledge (i) by providing an effective open-source tool to uncover untrustworthy website networks based on the utilization of the same Google Analytics/AdSense IDs, with the added ability to expose networks based on historical data, and (ii) by providing insight into the Slovak untrustworthy website landscape through delivering a first of its kind mapping of Slovak untrustworthy website networks. Our approach is based on a mix-method design employing a qualitative exploration of data collected in a two-wave study conducted in 2019 and 2021, utilizing a custom-coded tool to uncover website connections. Overall, the study succeeds in exposing multiple novel website ties. Our findings indicate that while some untrustworthy website networks have been found to operate in the Slovak infosphere, most researched websites appear to be run by multiple mutually unconnected administrators. The resulting data also demonstrates that untrustworthy Slovak websites display a high content diversity in terms of connected websites, ranging from websites of local NGOs, an e-shop selling underwear to a matchmaking portal.

## 1 Introduction

The burgeoning of misleading or false information in cyberspace has emerged as a central concern on the public agenda in recent years [1], fueled by, among other things, the sprouting of untrustworthy websites across many countries around the world, with Slovakia being no exception [2]. Notwithstanding that these untrustworthy websites [3], in some instances, have been ascribed to exercise far-reaching influence over the public discourse [1, 4], transparency and accountability of such websites remain an issue in most cases [5], with multiple websites remaining shrouded in anonymity. A useful approach that has crystallized in data journalism to uncover connections among websites and map untrustworthy website networks [6], has been a technique utilizing third-party tracking IDs mainly used by web analytics and ad serving services.

Web analytics is an indispensable tool for any website to understand and optimize its web usage [7], with a significant number of website administrators opting for a freemium third-party service such as Google Analytics [8]. Given that the client sides of such third-party services report to the same centralized backend, the client requests need to contain distinctive identifiers for the backend servers to differentiate between clients. However, such a setup is not exclusive to technologies tracking web usage but is also present in various other third-party services, such as the widely utilized ad serving service AdSense. The utilization of these third-party services among Slovak and Czech untrustworthy websites is high, with the most popular analytics technology being Google Analytics [9]. For convenience and in some cases also per the services' applicable terms and conditions [10], website administrators often manage their websites under a single identifier (ID). As a side result of using a single ID across multiple websites, it is possible to uncover connections between domains that otherwise seem unconnected. As this tracking code technique has garnered wider attention, it is possible that website administrators might have dissociated their IDs across their websites, which in return poses threats to the current widely utilized approaches aimed at uncovering such networks.

Notwithstanding that the threat posed by untrustworthy websites is a central concern on the public agenda [1], published work on untrustworthy website network detection remains limited, with work on this topic largely confined to reports and articles by non-governmental organizations, enthusiastic code developers and investigative journalists [11, 12]. As for the work exploring untrustworthy website networks in Slovakia, the inquiry remains even more so limited, with only a few investigative reporting pieces by local journalists [13, 14]. Thus, the aim of this paper is two-fold, (i) to provide an effective open-source tool to uncover untrustworthy website networks utilizing the same Google Analytics and AdSense IDs, with the added ability to expose networks based on historically associated IDs; and (ii) to provide insight into the Slovak untrustworthy website landscape by delivering a first of its kind mapping of Slovak untrustworthy website networks. To achieve the above aim, the presented article first offers a short state of the filed overview as well as background information on the untrustworthy websites landscape in Slovakia. Then, it introduces the methods and results of our empirical study, utilizing a highly effective custom python script used to uncover website connections in a longitudinal study of untrustworthy Slovak websites, with data collection undertaken first in April/May 2019 and then in April/May 2021. In the results section of this article, we then also present a short qualitative description of two untrustworthy Slovak website networks. Finally, based on the findings of our study, at the end of this article, we provide a brief discussion highlighting limitations as well as charting possible avenues for future research.

## 1.1 Related work

The utilization of website tracking codes to uncover hidden connections between websites was likely first reported by Baio in 2011 [15]. However, the technique garnered wider attention in the disinformation research community mostly after a social media analyst named Lawrence Alexander published in 2015 an investigative report in which he utilized Google Analytics IDs to uncover a pro-Kremlin web campaign [11]. However, the published approach was laborious and hardly scalable as it heavily relied on a manual component. This laboriousness prompted Seitz and Alexander to release a computer script in 2015 [16], which automated some of the tasks. The script was in 2017 updated as a number of the backend services it relied upon closed down [17]. It is around this time that also some academic scholarship on this topic emerged [7]. The utilization of Google Analytics and Google AdSense IDs to uncover untrustworthy website networks has become nowadays a staple in data journalism, however, with journalists still oftentimes relying on manual data collection [6]. As the tracking code linking technique has become better know also among the wider public, it might be argued that untrustworthy website administrators might have taken steps to dissociate their Analytics and AdSense IDs across their websites, which in return poses threats to the current publicly available scripts aimed at uncovering such networks [17]. As for the work exploring untrustworthy website networks in Slovakia, the inquiry remains scarce, confined mainly to a couple of traditional qualitatively based investigative journalism pieces, notably by Benčík [18] and Šnídl [19], and investigative data journalism pieces notably by Breiner [13, 14]. With the latter one mentioned succeeding in uncovering two higher-profile untrustworthy website networks, both spreading predominantly false health claims [13, 14].

## 1.2 The Slovak untrustworthy website landscape

Untrustworthy websites have in recent years started to burgeon across many countries around the world, with Slovakia being no exception to this practice. However, the exact number of such websites, including in Slovakia, is difficult to pinpoint. This situation is, among other things, fueled by the fact that websites can, in general, get easily taken down or redirected to a different domain, while new websites can, due to relatively low entrance barriers, be created also reasonably quickly [2]. As for the Slovak untrustworthy websites landscape, the most comprehensive publicly available list of such websites is compiled by a local media watchdog named Konšpirátori.sk [20]. The list, which is heavily relied upon by local researchers and policy experts [2, 21], ranks websites on a ten-point scale. Those scoring more than six points are considered to have highly dubious, deceptive, fraudulent, conspiratorial or propaganda content [22]. As of the end of May 2021, konspiratori.sk lists 210 websites with a score higher than six [20]. Thematically the websites tracked in the konspiratori.sk list cover a wide range of topics, including, among others, health disinformation, Russian propaganda, and the paranormal.

According to website traffic data estimates by SimilarWeb, some of the Slovak untrustworthy websites boast a high number of monthly visitors, with some of these websites even ranking in the top 100 most visited websites in Slovakia [2, 9]. However, transparency remains an issue in most of the cases. Multiple websites were shown to be actively trying to conceal their identity by, among other things, utilizing various domain privacy services or offshore hosting [5]. From a financial perspective, untrustworthy Slovak websites rely on various business models to sustain operation, with popular income sources including tax designation, e-commerce, crowdfunding, and advertising [5].

## 2  Methods

To study the networks of untrustworthy Slovak websites, we used a mix-method approached employing a qualitative exploration of data collected from a custom coded python script tasked to uncover websites using the same Google Analytics/AdSense identifier. The research was approved by the Matej Bel University (UMB) Ethics Committee (Reference no.: 1113-2017-FF). All methods were performed in accordance with UMB ethics guidance and regulations. Data Collection was conducted in two waves, first in April/May 2019 and then in April/May 2021. For the April/May 2019 data collection phase, our initial untrustworthy websites list consisted of 144 websites, available at that time, taken from the konspiratori.sk database. For the April/May 2021 data collection phase, our initial untrustworthy websites list consisted of 205 websites, also taken from the konspiratori.sk database [20]. Inclusion of a website in the database is based on a set of publicly available criteria, against which a Review Board consisting of, inter alia, prominent historians, political scientists, medical professionals, journalists and civil society representatives assess a website [23]. In utilizing the konspiratori.sk database, we followed best practices used in Slovak disinformation scholarship [2, 5, 21]. Second, each website in our dataset was during the respective data collection phases manually checked to determine its availability status and primary language. Unavailable websites and websites with a primary language other than Slovak were discarded. For the data collection phases, we build a python script, coded in python 2.7.10. The script builds upon code by Alexander and Seitz [16] and Seitz [17]. However, as opposed to their script, our script is based on a modified data collection setup designed to mitigate irrelevant data ballast and filter out data relics from third-party services that it utilizes, such as the Spy on Web API. In addition, our script also incorporates various novel features, most notably a history function, i.e., the ability to collect tracking codes from historical versions of websites, thus helping to evade possible measures undertaken by website administrators to dissociate their Google Analytics and AdSense IDs. For a simplified flowchart of the script, see Fig. 1.

The script underwent multiple rounds of testing and validation, including against manually collected data. The first phase of data collection using our custom python script was carried out during April/May 2019. After downloading the data, the script automatically linked websites based on the utilization of the same Google Analytics/AdSense identifier and created a graph file. Concurrently with the automatic data collection utilizing our script, we also performed a manual data collection, downloading the necessary data about all websites on our untrustworthy website list. The manually collected data represented a validation dataset against which the python-script-based data was compared. The second phase of data collection was carried out in two separate streams during April/May 2021. The first stream aimed to validate the history function of our script, i.e., the ability to collect tracking codes from historical versions of a website. The second stream represented a longitudinal continuation of the first data collection phase conducted in April/May 2019.

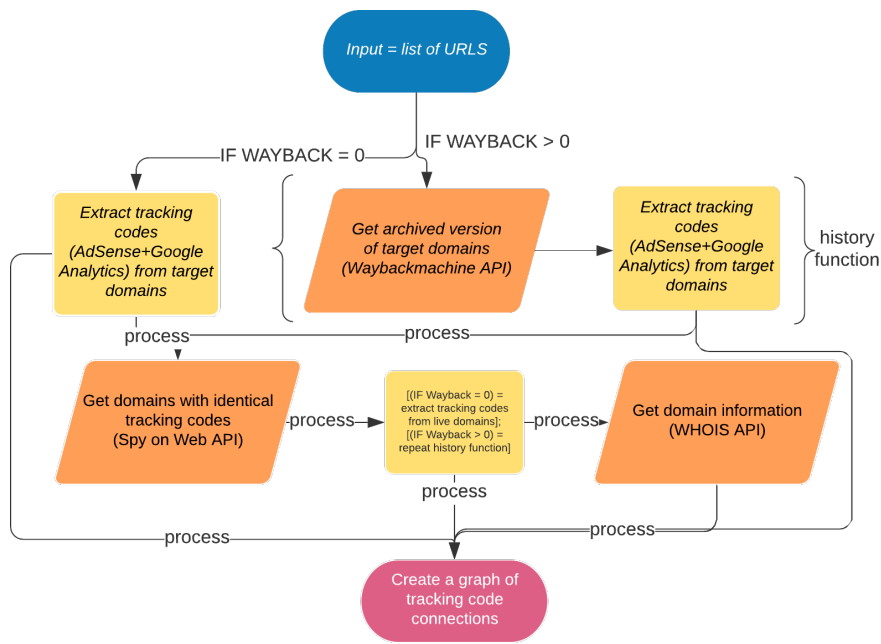

**Fig. 1.** Simplified flowchart of our custom unmasking script.

After each data collection phase ended, all datasets underwent data cleansing to detect and correct corrupt or inaccurate records. Subsequently, we calculated basic descriptive statistics for the cleansed datasets and identified the most salient disinformation networks. The resulting data from the two data collection phases were then content-coded and qualitatively examined, focusing mainly on the success rate of our approach in uncovering untrustworthy website networks. Last, we formatted the resultant datasets and visualized the data.

## 3 Results

Out of our initial sample size of 144 untrustworthy websites in the first data collection phase undertaken in April/May 2019, 21 websites appeared to cease to exist, and 74 were found to be primarily in languages other than Slovak, and therefore were discarded from our dataset. In addition, we also discarded three micro-blogging websites, as tracking codes present on such websites are managed by the respective platform operator, e.g., livejournal.com. Thus, our final list of untrustworthy websites for the first data collection phase consisted of 46 entries. As for the second data collection phase, out of our initial sample size of 205 untrustworthy websites, 51 appeared to cease to exist, and 87 were found to be primarily in languages other than Slovak and were therefore discarded from our dataset. In addition, two micro-blogging websites were discarded. Thus, our final list of untrustworthy websites for the second data collection phase consisted of 65 entries. A manual content-coding of the final lists of untrustworthy websites, utilizing a modified framework [9] based on the one developed by Mintal and Rusnak [5], showed that the majority of websites under investigation were in both data collection phases News-Focused; for a more detailed breakdown per each content category see Table 1.

**Table 1.** Frequency table of websites according to their content category.

| Content category | Frequency (April/May 2019 data) | Frequency (April/May 2021 data) |
| --- | --- | --- |
| News-Focused | 20 | 39 |
| Ideological or Supporting Cause | 9 | 13 |
| Health and Lifestyle | 13 | 12 |
| Paranormal | 4 | 1 |

After extensive code review, syntax check and debugging, our custom python script demonstrated a data retrieval accuracy of =100.00%, validated against four rounds of manually collected data from 20 randomly selected websites. Testing was conducted from December 2018 to February 2019. In addition, the data for all 46 untrustworthy websites on our list was concurrent with the automated data collection utilizing our script, also collected manually. A comparison between the automated and manually collected datasets, with the later one used as a validation dataset, revealed a data retrieval accuracy of =100%.

The data collected using the script during the April/May 2019 data collection phase showed that 84.79% of websites under investigation contained either a Google Analytics or Google AdSense tracking ID. Out of the 46 websites on our list, 14 websites were uncovered to belong to a network of linked websites, with such ecologies amounting to (n=11). The largest number of uncovered networks (n=6) were connected to News-related content-coded untrustworthy websites, (n=3) were connected to Health and Lifestyle websites, and (n=2) were connected to Ideological or supporting cause websites. The obtained data about the uncovered networks were during July/August 2019 qualitatively benchmarked against publicly available investigative pieces and data from the Business of Misinformation Project at CEU. The project employed a predominantly qualitative approach to uncover untrustworthy website ownership data in Slovakia, relying on investigative reports, the Investigative Dashboard Databases, self-reported data, and WHOIS data [5]. The benchmarking exercise yielded positive results, with our quantitative approach uncovering ten at that time unknown networks, with some of them internally classified as high-interest ones (e.g., pub-2531845767488846/UA-12857229; pub-9657897336906985; UA-1374898). The classification was based on external and internal consultations with relevant stakeholders. A qualitative analysis of the 12 uncovered networks showed a high diversity of linked websites in these networks, with websites belonging to a local NGO (e.g., UA-5743998), an e-shop selling underwear (UA-1374898), a matchmaking portal (pub-4883385023448719), but in some cases also to multiple other untrustworthy websites spreading dubious health claims (pub-9657897336906985). The descriptive statistics for the uncovered networks for both data collection phases are shown in Table 2.

For the second phase of data collection, set during April/May 2021, the script passed repeated code review and testing. The history function of the script, i.e., the ability to collected tracking codes from historical versions of a website, was validated against data collected in 2019, with a data retrieval accuracy of =100%. The data collected using the script during the second data collection phase revealed that 67.69% of websites under investigation contained either a Google Analytics or Google AdSense tracking ID. Out of the 65 websites on our untrustworthy website list, 12 websites were identified to belong to a network of connected websites, with such ecologies amounting to (n=11). The largest number of uncovered networks (n=7) were linked to News-related content-coded untrustworthy websites from our list, (n=2) were linked to Health and Lifestyle websites, and (n=2) were linked to Ideological or supporting cause websites. A qualitative analysis of the 11 uncovered networks again showed a high diversity of affiliated websites in these networks. Compared to the previously identified networks, one network appeared to cease to exist, and a new one was detected (UA-24461628). The descriptive statistics for the uncovered networks for both data collection phases are shown in Table 2.

**Table 2.** Descriptive statistics for uncovered networks.

|         | Dimension of networks (n=11) (April/May 2019 data) | Dimension of networks (n=11) (April/May 2021 data) |
|---------|---|---|
| Min—Max | 2—9   | 2—7   |
| Mean    | 3.636 | 3.181 |
| SD      | 1.919 | 1.465 |

When exploring the degree of change among the two datasets (2019 and 2021) in terms of websites modifying their Google Analytics ID, Google AdSense ID, or both — out of 46 websites tracked in the first data collection phase, one website has deleted its AdSense ID, one its Google Analytics ID, while three websites have added an additional ID type to their already used one. Comparing the degree of change for the uncovered linked websites is, however, higher, with multiple instances of websites deleting the IDs that connected them to one of the 46 websites on the konspiratori.sk list, with examples including the official website of a high-profile Slovak actor (UA-12857229), or a website of a high-profile Slovak singer (UA-1374898).

### 3.1 Sample description of two uncovered networks

As for a qualitative exploration of some of the uncovered networks, considering the nature and page limitations of this short study, we offer a brief case description of two thematically different networks, with the second one being publicized for the first time.

The first network concentrates around AdSense ID — pub-9657897336906985. The network comprises multiple untrustworthy websites spreading dubious health claims, with the network and its perpetrator being publicly disclosed in national media by a local data journalist in 2020 [13]. Breiner's discovery of the existence of such a network was undertaken independently from our 2019 findings, as the data from our first data collection wave were only disclosed on a need-to-know basis to certain stakeholders due to the nature of the information. However, contrary to our research, Breiner's investigative journalism piece also focused on uncovering the identity of the perpetrator — a young man from southern Slovakia [13], who appeared to be running the website network purely due to profit-making. These findings underscore the strong business-focused motives of some perpetrators operating untrustworthy websites. Such motives have also been highlighted by earlier research findings from the Business of misinformation project at CEU, which uncovered that 38 out of 49 researched untrustworthy Slovak websites operating in 2019 displayed ads or sold goods and services [5].

The second network, a previously unpublished one — pub-2531845767488846/UA-12857229 seems to be concentrated around a local IT services company, with a member of the company management shown in the past to directly back the establishment of a now-notorious untrustworthy website named panobcan.sk [24]. The network concentered around panobcan.sk' AdSense/Analytics ID is mainly interesting as in the past it used to be among others connected to a website of a high-profile Slovak actor, or that of *Slovenské národné noviny,* a newspaper officially published by *Matica slovenská* — a Slovak government-funded cultural and scientific institution primarily tasked with cultivating and presenting Slovak national cultural heritage [25]. While it is possible that the linkages among the observed websites stemmed from purely business-oriented grounds, as the company at the center of the website network focuses on providing IT services, at least in the case of the *Slovenské národné noviny,* a connection also appears to be on the content side, with researchers observing various dubious and at times also pro-Kremlin narratives spread in the past by the publisher of the newspaper — *Matica Slovenská [26].*

However, it should be noted that the presented motives or thematic narratives of the two described networks are in no way exhaustive, as the two networks were chosen based on internal and external consultations labelling them as high-interest ones. A narrative analysis of untrustworthy Slovak website networks would, however, represent an interesting avenue for future research.

## 4 Discussion

The results described above indicate a high data retrieval accuracy, precision, and efficiency of our script in uncovering untrustworthy Slovak website networks. The Slovak data obtained using our script show that while some untrustworthy website networks have been found to operate in the Slovak infosphere, most untrustworthy websites appear to be run by multiple mutually unconnected administrators. However, our data has also demonstrated that untrustworthy Slovak websites display a high content diversity of connected websites, ranging from websites of local NGOs to a matchmaking portal.

Our findings run somewhat counter to popular discourse and published investigations from other countries, in which large-scale foreign influence operations utilizing vast untrustworthy website networks administered from abroad have been discovered [27]. Our data on untrustworthy Slovak website networks rather indicates that they are largely domestically based.

However, the inhere presented approach of uncovering untrustworthy website networks is certainly not without limitations, which stem mainly from four interlinked sources. First, even though the utilization rate of Google Analytics and AdSense IDs is generally high, especially among the inhere researched Slovak websites, not all untrustworthy websites use these services. Second, it has to be noted that while the usage of the same ID across multiple websites administered by the same publisher is highly convenient and, in the case of AdSense, also mandated as per the services' Terms and conditions [10], publishers can still theoretically find ways to circumvent this requirement. However, such bypassing would require ample resources, as AdSense technically prohibits accounts with duplicate payee records, which get verified using the user's bank details, and in some countries also via an SMS code, or a physically mailed out PIN code. The third source of limitation is more of a technical nature as due to the vast and dynamically growing number of websites in existence, it cannot be ruled out that not all websites with the same ID have been yet indexed or recently crawled by the third-party services that our script relies on. However, owing to the setup of the utilized services, it is likely that such possibly occurring websites are assumably either very recently established, lower traffic or both. The fourth source of limitation of this study stems from the fact that some untrustworthy websites might potentially be connected through financially or otherwise related parties outside of the digital realm, such as relatives, business partners, shell companies, and others. Hence, although the benchmarking exercise of our approach indicates high precision and efficiency in uncovering untrustworthy Slovak website networks, this does not mean that our approach is able to uncover all connections of untrustworthy websites. Steaming from the discussed limitations and our results, we consider several avenues for future work. First, given among others, the possible differences in utilization rates of Google Analytics and AdSense across different websites leaves room for future work on modifying our script to extend its functionality and be able to link websites also based on other types of tracking IDs. Moreover, as other untrustworthy networks get discovered in the future, the collected data opens up, among others, possibilities for better understanding the motives and behaviour of the perpetrators administering such websites.

In conclusion, besides providing an effective open-source tool to uncover untrustworthy website networks, with the added ability to expose networks based on historical data, our article also delivered a first of its kind mapping of untrustworthy Slovak website networks, showing among others that untrustworthy Slovak websites appear to be run by multiple mutually unconnected administrators. Overall, we hope that our open-source script and this study's results will assist not only various actors combating online disinformation but also encourage further work on uncovering and understanding untrustworthy website networks and the possible ways of addressing issues related to this vital topic.

**Funding and declaration of conflicting interest**
This work was supported by IBM (IBM Faculty Award 2017), VEGA (grant project n. 1/0433/18) and APVV (grant project n. APVV-20-0334). The authors declared no potential conflicts of interest with respect to the research, authorship, and/or publication of this article.

**Contributions**
J.M.M. designed the study; M.K. coded the python script with input from J.M.M.; K.F. supervised the research and provided funding; J.M.M. collected the data and analyzed the data with input from K.F.; J.M.M. wrote the paper with input from K.F. All authors reviewed the results and approved the final version of the manuscript.

**Data Availability**
The code is available via GitHub at doi:10.5281/zenodo.4783300; the code is published under a GNU General Public License. The data that support the findings of this study have been deposited in Zenodo at doi:10.5281/zenodo.4783287; the data are available upon reasonable request.